\documentclass[%        	Class options:
aps,%                   	American Physical Society
prc,%                   	Physical Review D
superscriptaddress,%    	Authors' addresses linked with superscripts
nofootinbib,%           	Does not treat footnotes as references
floatfix]%              	Fixes float errors
{revtex4}%              	REVTEX 4 Package used
\usepackage{graphicx,%  	Default Latex 2eps package for embedding figures
longtable,color}%             	Useful for long table
\begin{document}
%======================================================================================
\title{Degeneracies of particle and nuclear physics uncertainties in neutrinoless $\beta\beta$ decay}
%--------------------------------------------------------------------------------------
%
%
%
\author{        E.~Lisi}
\affiliation{   Istituto Nazionale di Fisica Nucleare, Sezione di Bari, %\\
               Via Orabona 4, 70126 Bari, Italy}
\author{        A.M.~Rotunno}
\affiliation{   Dipartimento Interateneo di Fisica ``Michelangelo Merlin,'' %\\
                Via Amendola 173, 70126 Bari, Italy}%
\author{        F.~\v{S}imkovic}
\affiliation{	Department of Nuclear Physics and Biophysics,
				Comenius University, %\\
				Mlynsk\'a dolina F1, SK--842 15 Bratislava, Slovakia}
\affiliation{	Bogoliubov Laboratory of Theoretical Physics, JINR, %\\
				141980 Dubna, Moscow Region, Russia}
\affiliation{  Czech Technical University in Prague,
   				CZ-12800 Prague, Czech Republic}
\begin{abstract}%.......................................................................
\vspace*{0cm}
Theoretical estimates for the half life of neutrinoless double beta decay ($0\nu\beta\beta$) in candidate nuclei  
are affected by both particle and nuclear physics uncertainties, which may complicate 
the interpretation of decay signals or limits.  We study such uncertainties and their degeneracies in the
following context: three $0\nu\beta\beta$ nuclei of great interest for large-scale experiments 
($^{76}$Ge, $^{130}$Te, $^{136}$Xe), two representative particle physics mechanisms (light and heavy Majorana neutrino
exchange), and a large set of nuclear matrix elements (NME), computed within the quasiparticle random 
phase approximation (QRPA). It turns out that the main theoretical uncertainties, 
associated with the  effective axial coupling $g_A$  and with the nucleon-nucleon potential, 
can be parametrized in terms of NME rescaling factors, up to small residuals.
From this parametrization, the following QRPA features emerge: (1) the NME dependence on $g_A$ 
is  milder than quadratic; (2) in each of the
two mechanisms, the relevant lepton number violating parameter is largely degenerate with 
the NME rescaling factors; and (3) the light and heavy
neutrino exchange mechanisms are basically degenerate in the above three nuclei. We comment on the challenging theoretical and experimental 
improvements required to reduce such particle and nuclear physics uncertainties and their 
degeneracies.  
\end{abstract}%.........................................................................
%\medskip
%\pacs{%PACS Numbers:
%23.40.-s, 21.60.Jz, 02.50.-r} 
\maketitle

%%%%%%%%%%%%%%%%%%%%%%%%%%%%%%%%%%%%%%%%%%%%%%%%%%%%%%%%%%%%%%%%%%%%%%%%%%%%
%%%%%%%%%%%%%%%%%%%%%%%%%%%%%%%%%%%%%%%%%%%%%%%%%%%%%%%%%%%%%%%%%%%%%%%%%%%%
\section{Introduction}
%%%%%%%%%%%%%%%%%%%%%%%%%%%%%%%%%%%%%%%%%%%%%%%%%%%%%%%%%%%%%%%%%%%%%%%%%%%%
%%%%%%%%%%%%%%%%%%%%%%%%%%%%%%%%%%%%%%%%%%%%%%%%%%%%%%%%%%%%%%%%%%%%%%%%%%%%

The search for the neutrinoless mode of double beta decay ($0\nu\beta\beta$) in different $(Z,\,A)$ candidate nuclei, 
%...................................................
\begin{equation}
(Z,\, A)\to (Z+2,\, A) + 2e^- \ ,
\end{equation}
%...................................................
represents a major research program in experimental neutrino physics \cite{PDGR,Sc12,El12,Cr14,Go15}. From the theoretical viewpoint,
this decay mode provides a unique probe of the Dirac or Majorana nature of neutrinos \cite{Petc} and, in general, of lepton number violation (LNV) processes \cite{Ro11,Ve12,De12,Bi15}.   Indeed, the decay may be mediated not only by Majorana neutrinos with light (sub-eV) masses, but also by other LNV mechanisms involving new particle physics at higher mass or energy scales \cite{Petc,Ro11,Ve12,De12,Bi15}.

For a single LNV mechanism (labeled by the index $j$) occurring in a candidate nucleus (labeled by the index $i$), 
the $0\nu\beta\beta$ decay half life $T_i$ can be expressed as
%...................................................
\begin{equation}
\label{Ti}
T_i^{-1} = G_i^j\, |M_i^j|^2\, (\lambda^j)^2\ ,
\label{main}
\end{equation}
%...................................................
where $G_i^j$ is a kinematical phase-space factor, $M_i^j$ is the nuclear matrix element (NME) encoding the
nuclear dynamics of the process, and $\lambda^j$ is a LNV parameter encoding the particle physics aspects of decay.
 In principle, by means of independent $0\nu\beta\beta$ decay observations in different nuclei, 
one may hope to disentangle the underlying particle physics mechanism and the associated $\lambda^j$ value 
\cite{De07,Ge07,Fo09,Si10,Fa11,Al11}. 
Unfortunately, the relatively large nuclear model uncertainties affecting $M_i^j$ \cite{Fa09,Fa13,En15}
make this program quite difficult in practice.

Several studies have investigated the stringent conditions on the $M_i^j$ uncertainties,  under which various particle
physics mechanisms may---or may not---be disentangled with future, multi-isotope $0\nu\beta\beta$ decay data 
(see \cite{De07,Ge07,Fo09,Si10,Fa11,Al11,Ro11,Rotu,Mi12,Ho13,Me13,Pa14,Dev1} for an incomplete bibliography). A new twist in this field has been provided by recent discussions on  
the axial coupling $g_A$, which could be significantly
suppressed with respect to the vacuum value $g_A^{\text{vac}}\simeq 1.27$,
due to nuclear medium and other ``quenching'' effects. In particular, $g_A$ values in the
reference range often used in the past literature (see, e.g., \cite{Os92,Vo12}),
%------------------------
\begin{equation}
1\lesssim g_A\lesssim 1.27\ , 
\label{range}
\end{equation}
%----------------------
are being increasingly questioned by phenomenological studies. In fact,
in the framework of the quasiparticle random  phase approximation (QRPA), best-fit values 
$g_A<1$ were obtained in \cite{Fa08} by a joint analysis of the experimental data available for
the electron capture (EC), single beta ($\beta$) and two-neutrino double beta ($2\nu2\beta$) processes in the $^{100}$Mo and $^{116}$Cd nuclei.

 For the same weak processes and nuclei, such early indications for $g_A<1$ \cite{Fa08} 
   were later confirmed in another QRPA approach \cite{Su13}. 
Recently, a marked preference for values of $g_A$ below unity has also been found in a
different approach based on the interacting boson model (IBM)  \cite{Ba13,Yo13,Ba15}, where a possible 
dependence of $g_A$ on the atomic number has been noted \cite{Ba13}. See also \cite{Me11} 
for a recent discussion of quenching variations in the chiral effective field theory approach.
In the absence of a deep theoretical understanding of
quenching effects \cite{El12,Vo12}, these findings suggest that the usual reference range in Eq.~(\ref{range}) should
be conservatively extended somewhat below unity, e.g., in the range $0.8 \lesssim g_A\lesssim 1.27$ \cite{Ro13}.
%------------------------
%\begin{equation}
%0.8 \lesssim g_A\lesssim 1.27\ , 
%\label{range2}
%\end{equation}
%----------------------

In general, small (or uncertain) values of $g_A$ may strongly affect the half-life estimates via   
$T_i^{-1}\propto |M_i^j|^2 \propto g_A^4$, thus making even more difficult to constrain the particle physics 
mechanism and its LNV parameter \cite{Or14,Vi14}.  However, in the adopted QRPA framework \cite{Ro06},  
the $g_A$-dependence of $|M_i^j|$ is known to be {\em milder\/} than quadratic, 
since $g_A$ variations may be partly traded by shifts of another free parameter---the particle-particle strength $g_{pp}$ \cite{Si11}---via a joint fit to reference $2\nu\beta\beta$ data; see, e.g., \cite{Fa08,Vo12,Ro13,Si11,En14}.
Therefore, it makes sense to revisit the problem of determining the particle physics mechanism and its LNV parameter, by 
allowing $g_A<1$ within the QRPA. 

To this purpose, we
consider in the following three nuclei ($^{76}$Ge, $^{130}$Te, $^{136}$Xe) and two $0\nu\beta\beta$ mechanisms 
mediated by light ($L$) and heavy ($H$) Majorana neutrino exchange, within an updated QRPA approach.  
We show that, even if the decay half lives were
accurately measured, the current nuclear model uncertainties (mainly related to $g_A$ and to the nucleon-nucleon potential) 
would lead to a degeneracy between the LNV parameter and the NME errors in each mechanism,
as well as between the mechanisms themselves. Although 
limited to a few representative nuclei and  $0\nu\beta\beta$ decay processes, these results highlight 
the severe conditions and the challenging improvements needed to (partially) lift such 
degeneracies in the future.

Our work is structured as follows. In Sec.~II we introduce the notation and conventions for the two ($L$ and $H$) decay mechanisms. In Sec.~III we discuss the QRPA calculation of the nuclear matrix elements and the parametrization of the associated uncertainties. In Sec.~IV we perform a statistical analysis of prospective data, showing the degeneracy of particle and nuclear physics uncertainties,
both within each mechanism and between the two mechanisms. We briefly summarize our results in Sec.~V.
 
\vspace*{-2mm}
%%%%%%%%%%%%%%%%%%%%%%%%%%%%%%%%%%%%%%%%%%%%%%%%%%%%%%%%%%%%%%%%%%%%%%%%%%%%
%%%%%%%%%%%%%%%%%%%%%%%%%%%%%%%%%%%%%%%%%%%%%%%%%%%%%%%%%%%%%%%%%%%%%%%%%%%%
\section{Light and Heavy Neutrino Exchange: Notation and conventions \label{SecII}}
%%%%%%%%%%%%%%%%%%%%%%%%%%%%%%%%%%%%%%%%%%%%%%%%%%%%%%%%%%%%%%%%%%%%%%%%%%%%
%%%%%%%%%%%%%%%%%%%%%%%%%%%%%%%%%%%%%%%%%%%%%%%%%%%%%%%%%%%%%%%%%%%%%%%%%%%%

In the following, we shall study $0\nu\beta\beta$ decay in three representative nuclei of great interest for large-mass projects,
%------------------
\begin{equation}
i=1,\,2,\,3 =  {}^{76}\text{Ge},\, {}^{130}\text{Te},\, {}^{136}\text{Xe}\ ,
\label{indexi}
\end{equation}
%----------------------- 
and two reference LNV mechanisms, mediated by either light ($L$)
or heavy ($H$) Majorana neutrino exchange \cite{Ha83,Pa99},
%------------------
\begin{equation}
j=1,\,2 =  L, \,H\ .
\label{indexj}
\end{equation}
%----------------------- 
We refer the reader to \cite{Ro11,Ve12,De12,Bi15} for recent discussions of the particle physics dynamics of the $L$ and $H$ mechanisms, and to \cite{Go14} for
an approach interpolating between these two cases.

Here we just remind that, for the $L$ mechanism, the LNV parameter can be expressed as \cite{PDGR}:
%-------------------------
\begin{equation}
\lambda^L = m_{\beta\beta} = \left| \sum_{h=1}^3 |U_{eh}|^2 e^{i\phi_h} m_h \right|\ ,
\label{LNVL}
\end{equation}
%-----------------------
where $m_h$ are the masses of the three known light neutrinos $\nu_i$, $U_{eh}$ are their mixing matrix elements
with $\nu_e$, and $\phi_h$ are unconstrained Majorana phases (one of which can be rotated away).
For the $H$ mechanism, the LNV parameter can be expressed as (see, e.g., \cite{Pa99}):
%-------------------------
\begin{equation}
\lambda^H = M_{\beta\beta} = m_e \left| \sum_{k\geq 4} |U_{ek}|^2 e^{i\Phi_k} \frac{m_p}{M_k} \right|\ ,
\label{LNVH}
\end{equation}
%-----------------------
where  $M_k$ are the masses of possible heavy neutrinos beyond the known ones 
(assuming $M_k\gg m_p$), $U_{ek}$ are their mixing matrix elements
with $\nu_e$, and $\Phi_k$ are further Majorana phases. 

The $L$ and $H$ mechanisms are characterized by the same phase space \cite{Pa99}, 
%------------------
\begin{equation}
G_i^L=G_i^H\equiv G_i\ ,
\end{equation}
%----------------------- 
which, in the conventions of \cite{Ro06}, embeds  a factor $1/m_e^2$, so that its units are 
$[G_i]=y^{-1}\,\text{eV}^{-2}$, while $[\lambda^j]=\text{eV}$.  

Finally, we linearize Eq.~(\ref{main}) by taking logarithms  as in 
\cite{Fo09,Fa09,Rotu}, 
%-----------------
\begin{equation}
\tau_i = \gamma_i -2\eta_i^j - 2\mu^j\ ,
\label{linear}
\end{equation}
%-------------------
where
%------------------------
\begin{eqnarray}
\tau_i &=& \log_{10}(T_i/\text{y})\ ,\label{tau}\\ 
-\gamma_i &=& \log_{10}[G_i/(y^{-1}\,\text{eV}^{-2})]\ ,\label{gamma}\\
\eta_i^j &=& \log_{10}|M_i^j|\ ,\label{eta}\\
\mu^j &=& \log_{10}({\lambda^j/\text{eV}})\label{mu}\ .
\end{eqnarray}
%-------------------------

Table~I reports the numerical values of $G_i$  (and of $\gamma_i$) as taken from \cite{Ia12},
together with the most stringent experimental lower limits for 
the half lives $T_i$ as quoted by GERDA \cite{GERD} for $^{76}$Ge (in combination with IGEX \cite{IGEX} 
and HdM \cite{HdM0}), by CUORE \cite{CUOR} for $^{130}$Te (in combination 
with CUORICINO \cite{CUO2}), and by KamLAND-Zen \cite{KZEN} for $^{136}$Xe (in combination with EXO-200 \cite{EXXO}). Note that the
$G_i$ values of \cite{Ia12} have been rescaled by the fourth power of $1.27$, for consistency with our conventions 
[see Eq.~(\ref{nmep}) below].

%==============================================================================================================
\begin{table}[t]
\caption{Phase space values \protect\cite{Ia12} and 90\% C.L.\ half-life limits for the three nuclei considered in this work. See the text for details.}
\begin{ruledtabular}
\begin{tabular}{ccccl}
$i$ & $G_i$ ($y^{-1}$~eV$^{-2}$) & $\gamma_i$ & $T_i$ (y) & Experiments \\
\hline
%--------------------------------------------------------------------------------------------------------------
$^{76}$Ge	&	$2.21\times10^{-26}$	& 25.656	& $>3.0\times 10^{25}$	& GERDA~+~IGEX~+~HdM  \protect\cite{GERD} 	\\
$^{130}$Te	&	$1.33\times10^{-25}$	& 24.876	& $>4.0\times 10^{24}$	& CUORE-0~+~CUORICINO \protect\cite{CUOR}			\\
$^{136}$Xe	&	$1.36\times10^{-25}$	& 24.865	& $>3.4\times 10^{25}$	& KamLAND-Zen~+~EXO-200   \protect\cite{KZEN}
\end{tabular}
\end{ruledtabular}
\end{table}
%==============================================================================================================

 We emphasize that, although the $L$ and $H$ mechanisms share the same phase-space factor
$G_i$, their dynamics is quite different. The exchange potentials behave like $1/r$ and delta functions in the $L$ and $H$ cases, respectively, inducing significant differences in the multipole decomposition of the corresponding matrix elements, as well as in in their sensitivity to $g_A$ changes (not shown). Furthermore, the $H$ mechanism is more sensitive than the $L$ one to the choice of the nucleon-nucleon potential (see Sec.~III~B). Given these intrinsic dynamical differences, one does not expect a priori that the $L$ and $H$ mechanisms are phenomenologically degenerate, as they turn out to be.

%%%%%%%%%%%%%%%%%%%%%%%%%%%%%%%%%%%%%%%%%%%%%%%%%%%%%%%%%%%%%%%%%%%%%%%%%%%%
%%%%%%%%%%%%%%%%%%%%%%%%%%%%%%%%%%%%%%%%%%%%%%%%%%%%%%%%%%%%%%%%%%%%%%%%%%%%
\section{QRPA Nuclear Matrix Elements and uncertainties \label{SecIII}}
%%%%%%%%%%%%%%%%%%%%%%%%%%%%%%%%%%%%%%%%%%%%%%%%%%%%%%%%%%%%%%%%%%%%%%%%%%%%
%%%%%%%%%%%%%%%%%%%%%%%%%%%%%%%%%%%%%%%%%%%%%%%%%%%%%%%%%%%%%%%%%%%%%%%%%%%%

\subsection{Nuclear Matrix Elements}

The $0\nu\beta\beta$ nuclear matrix element $M=M^j_i$ consists of the Fermi (F),
Gamow-Teller (GT) and tensor (T) parts which, in the adopted conventions, read \cite{Ro06,Anat,SRCO}:
%-------------------------------------
\begin{equation}
M =  \left(\frac{g_A}{1.27}\right)^2
\left(
- \frac{M_\mathrm{F}}{(g_A)^2} + M_\mathrm{GT} - M^j_\mathrm{T}
\right)\ .
\label{nmep}
\end{equation}
%-------------------------------------
Note that the effective axial coupling $g_A$ enters not only in the prefactor and in the Fermi matrix element, but also 
in the calculation of the GT and tensor constituents, due
to a consideration of the nucleon weak-magnetism terms \cite{Pa99}.

In this work, six sets of nuclear matrix elements ${M}^j_i$ have been calculated for each
of the three nuclei in Eq.~(\ref{indexi}) and of the two mechanisms 
in Eq.~(\ref{indexj}). In particular, for both light and heavy neutrino mass mechanisms, the calculation is based
on the QRPA with partial restoration of isospin symmetry \cite{PARE}. 
The nuclear radius $R=r_0 A^{1/3}$ with $r_0=1.2$ fm is used. 
For each pair $(i,j)$, the NME set includes 18 variants, according to three different sizes of the single-particle space (small, intermediate, large), two different types of nucleon-nucleon interaction (charge-dependent (CD) Bonn and Argonne) \cite{SRCO}, and three different values of $g_A$,
%..............................
\begin{equation}
g_A= 1.27,\ 1.00,\ 0.80\ ,
\end{equation}
%.......................
which are representative of unquenched, quenched, and strongly quenched axial couplings. We remind that, for each calculation, the $g_{pp}  $ value is fixed by imposing that the corresponding (theoretical) $2\nu\beta\beta$ half-life equals the experimental one in each nucleus \cite{Ro06,Anat}. Summarizing, the following analysis is based on a total of 108  $0\nu\beta\beta$ matrix elements, calculated within a QRPA framework which reproduces three $2\nu\beta\beta$ half-lifes by construction. 

Figure~1 shows the scatter plots of these 108 NME values in logarithmic scale, which visualize the correlations between pairs of the $\eta^j_i$ parameters in Eq.~(\ref{eta}). NME variants are distinguished by different marker types: blue and red for $L$ and $H$ mechanisms; full and hollow for Argonne and CD-Bonn potentials; and squares, circles and triangles for $g_A=0.80$, 1.00 and 1.27, respectively. Single-particle space sizes are not explicitly distinguished (see also below). 
Note that, for graphical convenience, the $M^j_i$ values for the $L$ mechanism have been multiplied by a factor 100. This figure reveals a very strong correlation of the NME's, which is especially evident in the plane charted by $^{136}$Xe and $^{130}$Te. In practice, it appears that the theoretical uncertainties associated to NME variants can be largely absorbed by overall scaling factors.

The strong correlations emerging in Fig.~1 imply two different types of degeneracies 
between nuclear and particle physics aspects of the decay, at least in the $(i,\,j)$ sets considered herein:
(a) for a given decay mechanisms, the nuclear model uncertainties are degenerate with the LNV parameter $\lambda^j$ \cite{Fa09}; and (b) the two different mechanisms ($L$ and $H$) are largely degenerate with one another \cite{Rotu}. With respect to \cite{Fa09,Rotu}, we
sharpen these statements by using a convenient 
parameterization and statistical treatment of the NME uncertainties within the QRPA (including cases with $g_A<1$), as discussed below.

 A final remark is in order. Variations of the single-particle space size (from small to intermediate and large size) produce relatively small NME changes, partly orthogonal to the main degeneracy directions of Fig.~1. These changes do not reveal a specific pattern: e.g., it turns out that, for $^{76}$Ge, the NMEs are slightly low for intermediate space size (with respect to small or large sizes), while the opposite happens for $^{130}$Te and $^{136}$Xe; for the latter two nuclei, the NMEs values for small and large sizes somewhat differ, which they are rather close for $^{76}$Ge. In the absence of
a compelling pattern emerging from different single-particle space sizes, we omit their distinction in Fig.~1, and adopt
the conservative viewpoint that the associated NME variations are basically uncorrelated among the three nuclei.

%---------------------------------------------------------------------------
\begin{figure}[t]
%\vspace*{+1.0cm}
%\hspace*{-1cm}
\includegraphics[scale=0.64]{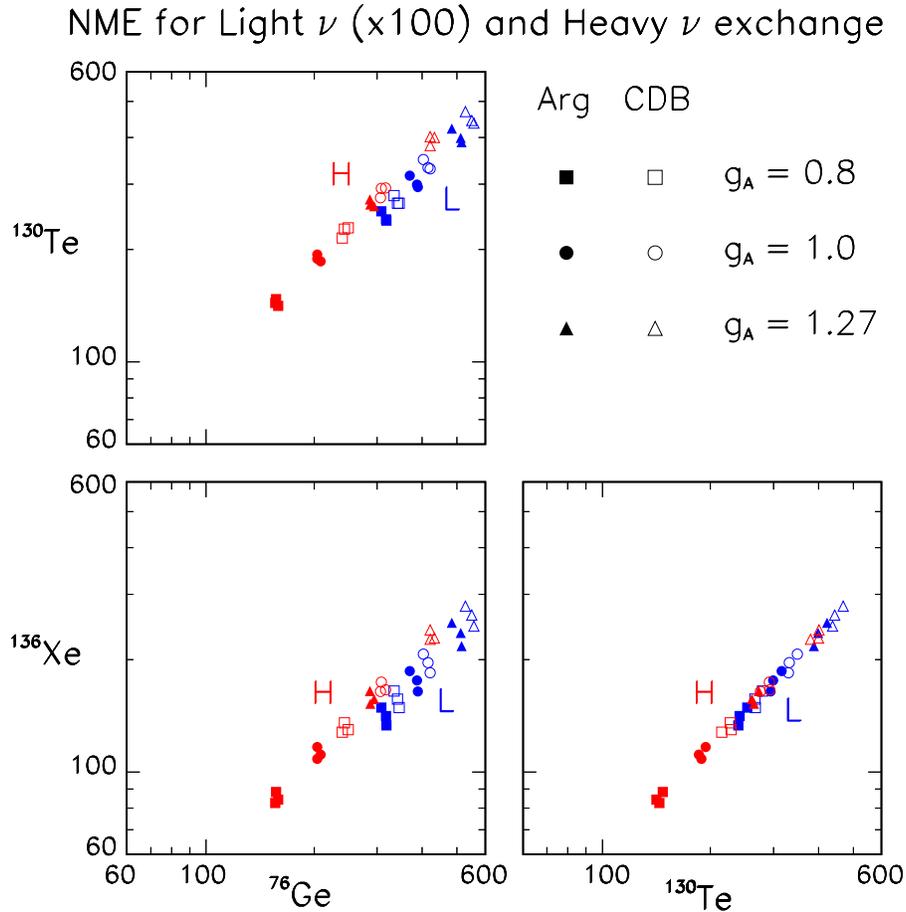}
%\vspace*{0.0cm}
\caption{\label{f01}
Scatter plot (in logarithmic scale) of the nuclear matrix elements $|M^j_i|$, for each pair of the three $0\nu\beta\beta$ candidate nuclei, and for both heavy ($H$) and light ($L$) Majorana neutrino exchange. In the latter case, the NME have been multiplied by a factor 100. See the text for details.}
\end{figure}
%---------------------------------------------------------------------------

\subsection{Parametrization of uncertainties}

We find that the spread of the numerical $\eta^j_i$ values can be fully covered by the following empirical parameterization, which incorporates both the strong linear correlation among the $\eta^j_i$ and the residual transverse scatter visible in Fig.~1:
%...............................
\begin{equation}
\eta^j_i = \bar \eta^j_i +\alpha^j (g_A-1)+s\beta^j\pm \sigma^j\ ,
\label{empiric}
\end{equation}
%...............................
where $s=+1$ ($-1$) for the CD-Bonn (Argonne) potential, and the parameters $\bar\eta^j_i$, $\alpha^j$, $\beta^j$ and $\sigma^j$ are given in Table~II. In the above equation, $\bar\eta_i^j$ represents a sort of ``central value'' for the set of $\eta_i^j$ values, while 
$\alpha^j(g_A-1)+s\beta^j$ represents the systematic theoretical bias due to admissible variations of $g_A$ with respect to the unit value, and of the nucleon-nucleon potential. In terms of NMEs ($|M| = 10^{\eta}$), the bias acts as an overall ($i$-independent) NME rescaling factor for the three nuclei. Finally, $\pm\sigma^j$ represents the residual range which is not covered by the previous bias, including variations due to the basis size (small, intermediate, large). [Actually, the $\sigma^j$ values covering the $\eta^j_i$ spread  depend slightly on the index $i$; we neglect these tiny variations, and conservatively take the largest value for $\sigma^j$.]
%------------

%\newpage
%==============================================================================================================
\begin{table}[t]
\caption{Numerical values for the empirical parametrization of nuclear model uncertainties in Eq.~(\protect\ref{empiric}), 
for the two $L$ and $H$ mechanisms.}
\begin{ruledtabular}
\begin{tabular}{ccccccc}
$j$ & $\bar\eta_1^j$ & $\bar\eta_2^j$ & $\bar\eta_3^j$ & $\alpha^j$  & $\beta^j$ & $\sigma^j$ \\ 
\hline
%--------------------------------------------------------------------------------------------------------------
$L$ & 0.600 & 0.504 & 0.267 & 0.458 & 0.021 & 0.032 \\ 
$H$ & 2.400 & 2.364 & 2.135 & 0.544 & 0.089 & 0.025 
\end{tabular}
\end{ruledtabular}
\end{table}
%==============================================================================================================

The above parametrization indicates that, within the QRPA, the functional dependence of the $0\nu\beta\beta$ NME on $g_A$ is significantly milder than the naive quadratic expectations ($|M^j_i|\propto g^2_A$), as
already noticed in \cite{Fa08,Si11,En14}. We recall that, within  the QRPA approach, the 
$g_{pp}$ parameter is adjusted to fit the $2\nu\beta\beta$ decay rate, and that both the $0\nu\beta\beta$ and $2\nu\beta\beta$ NME ($M^{0\nu}$ and $M^{2\nu}$) decrease
with  decreasing $g_A$ or with increasing $g_{pp}$. Then,  
if $g_A$ decreases, the $g_{pp}$ parameter must decrease as well, in order to 
keep $M^{2\nu}$ at the value fixed by the $2\nu\beta\beta$ half life $T^{2\nu}$, as shown in Fig.~2. As a consequence, also the change in the matrix element $M^{0\nu}$ is smaller than one might at first expect (see, e.g., \cite{Fa08}). In particular,  Eq.~(\ref{empiric}) suggests that, for $0\nu\beta\beta$ decay, the effective NME dependence
on $g_A$ is
close to be linear ($|M^j_i|\propto g_A$) rather than quadratic, at least for relatively small values of the difference $g_A-1$, and within the rough approximation $\alpha^j \sim 1/2 \sim \ln(10)$. In this sense,  the impact of the large $g_A$ uncertainties in the interpretation of $0\nu\beta\beta$ data \cite{Ba13,Or14,Vi14} may be effectively reduced (although not eliminated) within the QRPA approach  
\cite{Fa08,Vo12,Ro13}.

%---------------------------------------------------------------------------
\begin{figure}[bh]
\vspace*{0.6cm}
%\hspace*{-1cm}
\includegraphics[scale=0.38]{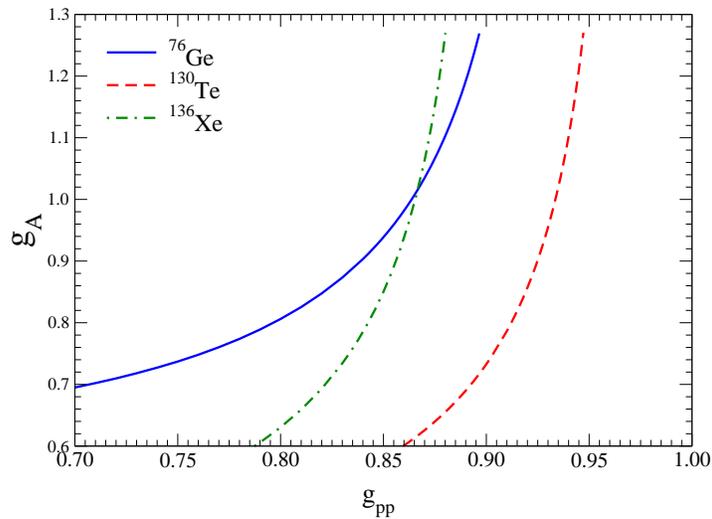}
%\vspace*{0.0cm}
\caption{\label{f02}
The relation between the weak axial-vector coupling parameter $g_A$ and the $g_{pp}$ parameter, 
determined from the measured $2\nu\beta\beta$  half-life of $^{76}$Ge, $^{130}$Te and $^{136}$Xe. 
The results refer to the case with Argonne potential and small size for the single-particle model space.
Similar results hold for other choices of the potential or model space (not shown).}
\end{figure}
%---------------------------------------------------------------------------

We remark that, for the sake of simplicity, $g_A$ has been assumed to be the same in the three considered nuclei. This seems to be an acceptable starting point for a phenomenological analysis, since the origin and amount of quenching are not well known. For instance, quenching effects might be assigned 
to the $\Delta$-isobar admixture in the nuclear wave function, or to the
shift of the GT strength to higher excitation energies due to short-range tensor correlations. Model-space truncation
can also exclude strength that may be pushed to high energies, and the omission of two-body currents can also leave 
excitations unaccounted for. Moreover, quenching effects might be different for different multipoles and, if associated 
with exchange currents, might be smaller for light nuclei. In the absence of a clear
picture for these effects, we have assumed that the same value of $g_A$ (either 0.8, 1.0, or 1.27) applies to all multipoles 
in the three medium-heavy nuclei consider herein. Of course, this simplified assumption may be revisited in future and more refined analyses of the quenching phenomenon within the $0\nu\beta\beta$ decay context.
In perspective, one should build a 
general theory of quenching in the nuclear medium, and
constrain systematically the theory with data from different nuclear processes linked to $0\nu\beta\beta$ decay,
so as to reduce the effects of the $g_A$ uncertainty (see, e.g., the
discussion in \cite{En15,Fr12}). At present, 
however, we must accept theoretical uncertainties  (at least) as large as in Eq.~(\ref{empiric}). 

 So far, we have mainly discussed the sensitivity to $g_A$ variations, characterized by the  $\alpha^j$ parameters. Let us now comment on the other parameters $\beta^j$ and $\sigma^j$. 
The value of $\beta^j$, which characterizes the NME sensitivity to the choice of the 
nucleon-nucleon potential, turns out to be much larger in the $H$ mechanism than in the $L$ one 
(by a factor of about four), as anticipated at the end of Sec.~II.
 Following the remarks at the end of Sec.~III~A, the theoretical uncertainty $\pm \sigma^j$ 
is treated as a ``one-standard-deviation range'', covering 
those residual uncertainties (including the single-particle space ones) which are not included in the ``bias'' term 
$\alpha^j(g_A-1)+s\beta^j$. This definition is conservative, because it allows
to cover [via Eq.~(\ref{empiric})] all the NME in Fig.~1, and not only 68\% of them. 

\subsection{Degeneracies in terms of observable half lives}

The results of Sec.~III~B allow to visualize the degeneracies mentioned at the end of Sec.~III~A in terms of observable quantities, i.e., the $0\nu\beta\beta$ decay half lives $T_i$ expected in different nuclei and for different underlying mechanisms. Since the $T_i$ are usually represented in logarithmic scale, we shall base our discussion directly on $\tau_i=\log_{10}(T_i/\mathrm{y})$. 
From Eqs.~(\ref{linear}) and (\ref{empiric}), the $\tau_i$ can be expressed as 
%............................
\begin{equation}
\tau_i = \gamma_i - 2\bar\eta^j_i - 2\bar\mu^j \pm 2\sigma^j\ ,
\label{corr}
\end{equation}
%..............................
where we have defined a ``rescaled'' LNV parameter $\bar\mu^j$ as,
%............................
\begin{equation}
\bar\mu^j = \mu^j +\alpha^j(g_A-1)+s\beta^j\ .
\label{deg1}
\end{equation}
%..............................
The above equations clearly show the degeneracy between the particle physics parameter $\mu^j$ 
and the systematic QRPA uncertainties parameterized by  $\alpha^j(g_A-1)+s\beta^j$: variations of the latter term can be traded for
opposite changes in the LNV parameter, without affecting the observable $\tau_i$.

Figure~3 shows the theoretical expectations in the planes charted by pairs of half lives $(\tau_k,\,\tau_h)$ 
in different nuclei, together with the current experimental lower limits as reported in Table~I. The half lives are 
correlated via Eq.~(\ref{corr}), which implies 
%............................
\begin{equation}
\tau_k-\tau_h = \gamma_k-\gamma_h-2(\bar\eta_k^j-\bar\eta_h^j)\ 
\label{diff}
\end{equation}
%...............................
for $j=H,\,L$, up to residual errors ($\pm2\sigma^j$), shown as crosses in Fig.~1. The position of each cross is irrelevant: the 
associated errors are the same at any point on the slanted lines---which should thus be thought as ``error bands''.  
This figure illustrates the second kind of degeneracy mentioned at the end of Sec.~III~B, namely, the near
indistinguishability of the $L$ and $H$ mechanisms, which is especially evident in the rightmost panel,
where the half-life expectations for the $L$ and $H$ cases are almost coincident. In the other two panels on the left,
the $L$ and $H$ lines are visually separated, but they can overlap within error bars. As far as the separation
between the slanted lines in Fig.~3 remains smaller or comparable than the theoretical error bars, future experimental data
on the half lives (no matter how accurate) will not be able to tell the $L$ from the $H$ mechanism.
In the following section, we shall quantify the two kinds of degeneracies by performing an analysis of prospective data.

%---------------------------------------------------------------------------
\begin{figure}[t]
%\vspace*{+1.0cm}
%\hspace*{-1cm}
\includegraphics[scale=0.69]{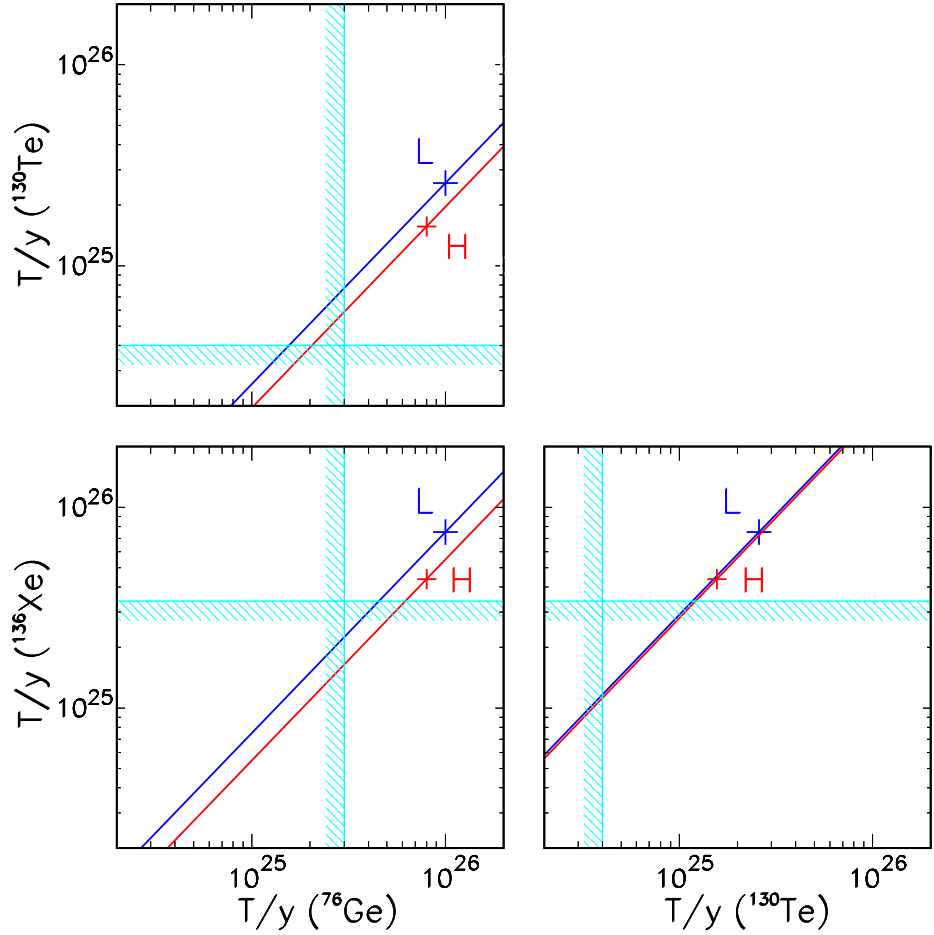}
%\vspace*{0.0cm}
\caption{\label{f03}
Correlation plot of expected half lives in different pairs of nuclei (slanted lines) together with residual theoretical errors (crosses) for the $L$ and $H$ decay mechanisms (blue and red lines, respectively). Also shown are the current experimental limits, as reported in Table~I. See the text for details.}
\end{figure}
%---------------------------------------------------------------------------

\section{Statistical analysis of prospective $0\nu\beta\beta$ data} 

There are good prospects to improve the current limits (Table~I) on the $0\nu\beta\beta$ half-life in the candidate nuclei ($^{76}$Ge, $^{130}$Te, $^{136}$Xe). In particular, in the next decade, upgrades of the current experiments or new planned projects should be able to explore an additional order of magnitude in $T_i$ \cite{El12,Sc12,Cr14,Go15} and hopefully find evidence for the decay. Below, we shall optimistically assume that a $0\nu\beta\beta$ decay signal is found in each of the three nuclei.

\subsection{Reference scenario}
 
We assume that the measured $^{76}$Ge half life takes a reference value of $10^{26}$~y. The half lives 
for $^{130}$Te and $^{136}$Xe in the $L$ and $H$ mechanisms are then obtained via Eq.~(\ref{diff}).
We also assume that each $T_i$ is measured within $\pm20\%$ at $1\sigma$,
corresponding to $\pm 0.08$ accuracy on $\tau_i$.  
Our prospective data sets are thus given by the following central
values and errors ($\bar\tau_i^j \pm s_i^j$) for the $L$ and $H$ cases, respectively: 
%...........................
\begin{equation}
\mathrm{data\ set\ }(L) \Leftrightarrow \left\{
\begin{array}{l}
\bar \tau_1^L\pm s_1^L = 26.000 \pm 0.080\ ,\\
\bar \tau_2^L\pm s_2^L = 25.412 \pm 0.080\ ,\\
\bar \tau_3^L\pm s_3^L = 25.875 \pm 0.080\ ,\\
\end{array}
\right.
\label{dataL}
\end{equation}
%............................
%...........................
\begin{equation}
\mathrm{data\ set\ }(H) \Leftrightarrow \left\{
\begin{array}{l}
\bar \tau_1^H\pm s_1^H = 26.000 \pm 0.080\ ,\\
\bar \tau_2^H\pm s_2^H = 25.292 \pm 0.080\ ,\\
\bar \tau_3^H\pm s_3^H = 25.739 \pm 0.080\ .\\
\end{array}
\right.
\label{dataH}
\end{equation}
%............................

%---------------------------------------------------------------------------
\begin{figure}[t]
%\vspace*{-0.8cm}
%\hspace*{-1cm}
\includegraphics[scale=0.69]{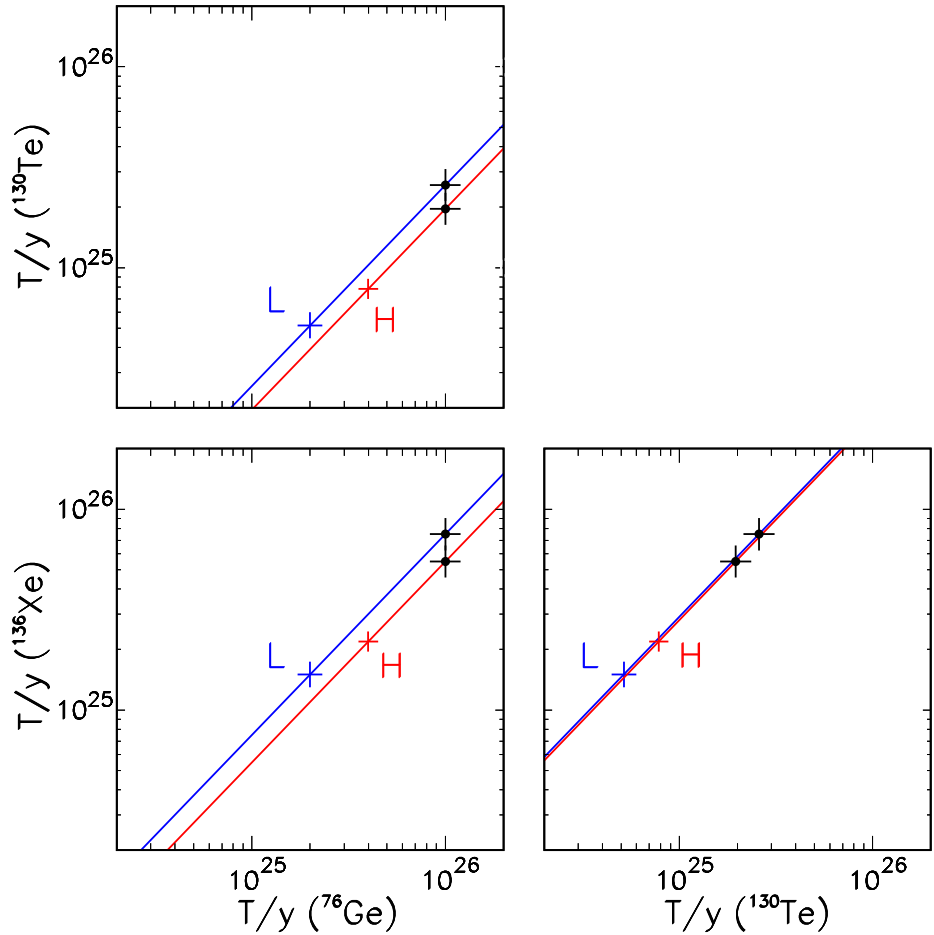}
%\vspace*{0.0cm}
\caption{\label{f04}
As in Fig.~3, but with lower limits replaced by prospective determinations of the half lives (black dots and crosses). See the text for details.}
\end{figure}
%---------------------------------------------------------------------------
Figure~4 shows the above data sets (black dots with crossed error bars) overlaid on the same theoretical predictions (colored slanted lines with errors) for the $L$ and $H$ cases as in Fig.~3. We discuss below the implications of these data on the degeneracy issues, by means of a statistical analysis. 

\subsection{Degeneracy between LNV parameter and QRPA uncertainties for fixed $0\nu\beta\beta$ decay mechanism}

Let us assume one of the two Majorana neutrino exchange mechanisms (either $L$ or $H$) as the ``true'' one for the $0\nu\beta\beta$ decay in the three nuclei. Given the previous QRPA calculations, the prospective data sets, and their associated uncertainties, we aim at determining the associated LNV parameter (either $\mu^{L}$ or $\mu^H$) and its errors. 

To this purpose, we consider the following $\chi^2$ function in terms of the ``rescaled'' parameter $\bar\mu^j$ in Eq.~(\ref{deg1}),
%....................
\begin{equation}
\chi^2(\bar\mu^j)=\sum_{i=1}^3\frac{(\tau_i-\bar\tau_i^j)^2}{(s_i^j)^2+(2\sigma^j)^2} = 
\sum_{i=1}^3\frac{(\gamma_i-2\bar\eta_i^j-2\bar\mu^j-\bar\tau_i)^2}{(s_i^j)^2+(2\sigma^j)^2}\ .
\label{chi2}
\end{equation}
%.................... 
Minimization of $\chi^2$ provides the central value $\bar \mu^j_c$ (at $\chi^2 = 0$, by construction) and its error $\delta^j$ (at $\Delta\chi^2=1$). We get:
%.......................
\begin{equation}
\bar\mu^j_c \pm \delta^j = \left\{
\begin{array}{l}
-0.772 \pm 0.029\ (L)\ ,\\ 
-2.572 \pm 0.027\ (H)\ .
\end{array}\right. 
\end{equation} 
%.......................  
The LNV parameter $\mu^j$ is then obtained from Eq.~(18) as 
%....................
\begin{equation}
\mu^j = \bar\mu^j_c -\alpha^j(g_A-1) - s\beta^j\pm \delta^j\ .
\label{LNVfit}
\end{equation}
%.................... 
The above expression for $\mu^j$ provides a useful breakdown of its uncertainties:  from
left to right, the terms following the central value $\bar\mu^j_c$ represent, respectively,
the systematic error due to $g_A$ variations from the unit value, the systematic bias 
due to the choice of the nucleon-nucleon potential (CD-Bonn {\em vs\/} Argonne), and the residual error
from theory and data uncertainties.

%---------------------------------------------------------------------------
\begin{figure}[t]
%\vspace*{-0.8cm}
%\hspace*{-1cm}
\includegraphics[scale=0.67]{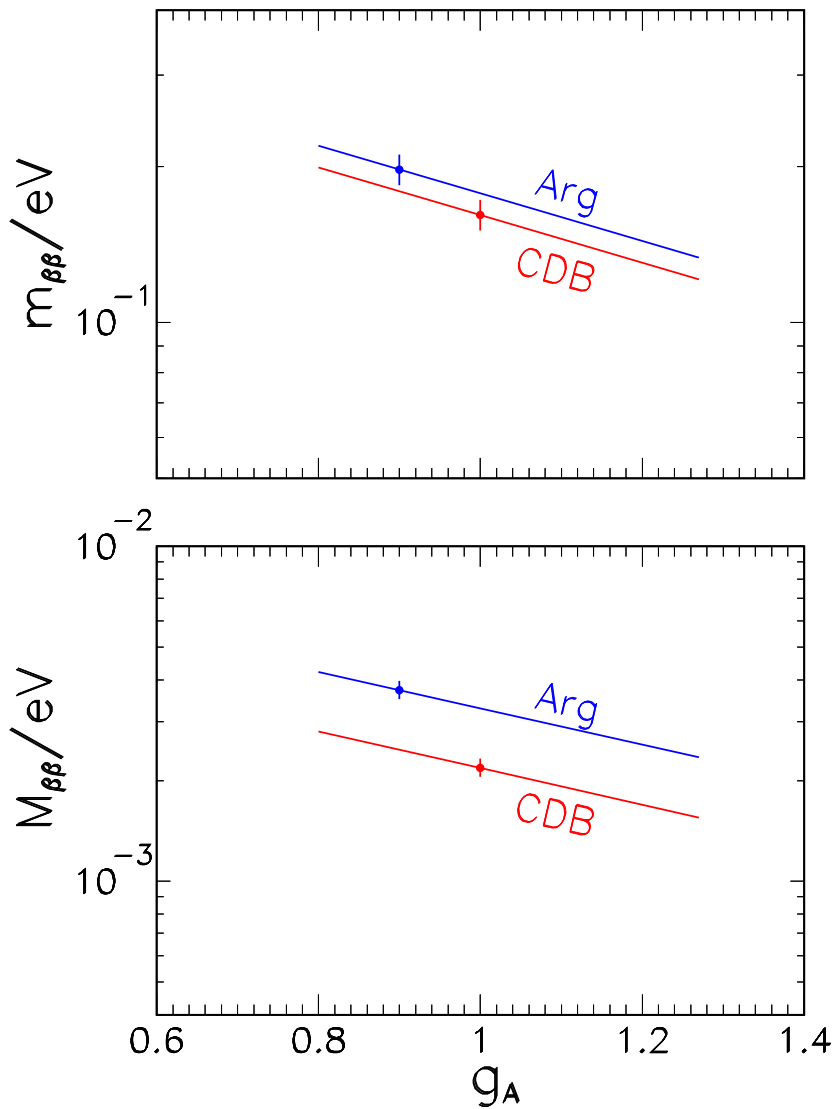}
%\vspace*{0.0cm}
\caption{\label{f05}
Upper panel: Light ($L$) Majorana neutrino exchange mechanism. The LNV parameters $m_{\beta\beta}$ (in eV), as derived from a fit to the prospective data in Eq.~(\protect\ref{dataL}), is shown as a function of $g_A$ (in the range $g_A\in [0.8,\,1.27]$).  The two curves refer to the CD-Bonn and Argonne choices for the nucleon-nucleon potential. The error bars attached to each line mark the size of the residual theoretical and experimental uncertainties from the fit. Lower panel: as above, but for heavy ($H$) Majorana neutrino exchange with LNV parameter $M_{\beta\beta}$, from a fit to the prospective data in Eq.~(\protect\ref{dataH}).}
\end{figure}
%---------------------------------------------------------------------------

Figure~5 represents the results of the above statistical analysis, in terms of the LNV parameters $m_{\beta\beta}$ 
 and $M_{\beta\beta}$ associated to the $L$ and $H$ mechanisms, as defined in Eqs.~(\ref{LNVL}) and (\ref{LNVH}), respectively. 
  The LNV parameters are shown (in logarithmic scale) as a function of $g_A$, whose variation in the representative range 
 $g_A\in [0.8,\,1.27]$ induces the main source of uncertainty: the higher $g_A$, the smaller $m_{\beta\beta}$ or
 $M_{\beta\beta}$. In decreasing order of relevance, the second source of uncertainty is represented by the nucleon-nucleon potential, whose twofold option  
 (CD-Bonn {\em vs\/} Argonne) splits the LNV parameter estimates into two curves. Finally, the third and smallest source of uncertainties 
 is induced by $\delta^j$ [see Eq.~(\ref{LNVfit})] and is shown as a representative error bar, on top of each curve. The total uncertainty affecting each LNV 
 parameter is given by the ``envelope'' of all these errors which, for $g_A\in [0.80,\,1.27]$, amounts to a factor of $\sim 2$
 for $m_{\beta\beta}$, and to a factor of $\sim 3$ for $M_{\beta\beta}$ (from minimum to maximum, in both cases).  
 
Overall uncertainties of a factor $\sim 2$ in the reconstructed value of $m_{\beta\beta}$ are definitely large, 
but not as dramatic as those of $O(10)$ discussed in \cite{Ba13,Or14}. This feature of the QRPA approach may thus be of
interest for relatively robust estimates of the experimental sensitivity to $m_{\beta\beta}$.  We stress that,
as  mentioned in Sec.~III~B, in the QRPA the 
effective dependence of the NME on the $g_A$ parameter  
($M^j_i\sim g_A$) is milder than in other frameworks (where $M^j_i\sim g_A^2$), as a result of the stabilizing  
role of the ``$2\nu\beta\beta$ calibration'' of the $g_{pp}$ parameter. 

In any case, 
limiting the $g_A$ range in Fig.~5, by means
of dedicated theoretical and experimental studies, will be a major step towards the reduction of 
the reconstructed LNV parameter uncertainties. One should also refine the understanding of
the nucleon-nucleon potential, so as to bring the two splitted curves in Fig.~5 closer to each other.
Such a long-term nuclear modeling program, although rather challenging, is warranted by the fundamental importance
of the worldwide  $0\nu\beta\beta$ decay search program.

%\vspace*{-1mm}
\subsection{Degeneracy between light and heavy neutrino exchange mechanisms for $0\nu\beta\beta$ decay}

As already mentioned, the ensembles of expected half lives  for the $L$ and $H$ mechanisms (slanted lines in Fig.~3) are very close
to each other, as compared to current theoretical uncertainties and prospective experimental errors (see Fig.~4).
This fact suggests that the two mechanisms are largely degenerate, i.e., they cannot be distinguished 
by data from the three candidate nuclei considered herein. One can quantify the degree of degeneracy as follows: 
the prospective data for one mechanism (say, $L$) are fitted with the predictions of the other mechanism (say, $H$), and 
viceversa. The value of $\chi^2_{\min}$ quantifies then the ``degree of misfit'': 
the lower is $\chi^2_{\min}$, the more difficult is to distinguish the $L$ and $H$ mechanisms. By using a $\chi^2$ approach
as in Eq.~(\ref{chi2}), we find that, as expected, the misfit is not statistically significant: $\chi^2_{\min}\simeq 1.1$ in
both cases. Therefore,
the two mechanisms are phenomenologically indistinguishable at the level of $(\chi^2_{\min})^{1/2} \sim 1\sigma$.

It should be noted that this small $1\sigma$ difference between the $L$ and $H$ scenarios 
is generated solely by the smallest sources of uncertainties (the data errors $s^j_i$ and the
residual theoretical errors $\sigma^j$), while it 
does not depend on the value of $g_A$ or on the choice of the nucleon-nucleon potential (which can be absorbed by variations of the unknown LNV parameter $\lambda_j$ in the fit). Improving  the latter two sources of
uncertainties, despite being crucial to assess $\lambda_j$, would not help to lift
the $L$-$H$ mechanism degeneracy. Breaking the degeneracy would require the challenging reduction of the non-parametric component
of the theoretical error $\sigma^j$ (which is at the level of $10^{\sigma_{j}}\simeq 5$-$7\%$ in our approach) and of the 
 prospective experimental errors $s_j^i$ (which we have assumed to be at the $\simeq 20\%$ level in this work). 

It is not obvious how the above stringent requirements can be achieved, even in the far future. In any case, we remind the reader that such conclusions refer to the specific decay mechanisms, candidate nuclei, and QRPA nuclear model considered in this work, and might 
thus be altered in a wider context. In general, favorable cases for the discrimination of any two decay mechanisms in a pair of candidate nuclei can be diagnosed, via correlation plots analogous  to our Figs.~1 and 3, 
by the emergence of a significant ``transverse'' separation of the slanted error bands (see also \cite{Rotu,Me13}). 

%\vspace*{-1mm}
\section{Summary}

We have studied in detail a phenomenological scenario involving three candidate nuclei for $0\nu\beta\beta$ decay
($^{76}$Ge, $^{130}$Te, $^{136}$Xe), two representative particle physics mechanisms (light and heavy Majorana neutrino
exchange, with LNV parameters $m_{\beta\beta}$ e $M_{\beta\beta}$), and a large set of nuclear matrix elements, 
computed within the quasiparticle random 
phase approximation. We have found that the main theoretical uncertainties, 
induced by the  effective axial coupling $g_A$  and with the nucleon-nucleon potential, 
can be parametrized in terms of rescaling factors for the nuclear matrix elements, up to small residuals.
Within the QRPA, the effective rescaling induced by $g_A$ variations is found to be almost linear, rather than 
quadratic in $g_A$ as naively expected. Despite this favorable feature, we find that, in each mechanism,
the relevant lepton number violation parameter is largely degenerate with the rescaling factors; in particular,
for $g_A\in [0.8,\,1.27]$ the total $m_{\beta\beta}$ ($M_{\beta\beta}$) uncertainty in numerical experiments amounts to a factor
of about two (three). Moreover, the light and heavy neutrino exchange mechanisms turn out 
to be largely indistinguishable from a phenomenological viewpoint. The stringent conditions needed to
lift the degeneracies between particle and nuclear physics aspects of $0\nu\beta\beta$ decay have been
briefly discussed. Progress may be envisaged, on the one hand, by studying
further decay mechanisms and candidate nuclei and, on the other hand, by understanding  
the various theoretical uncertainties associated to the QRPA and other nuclear models,
and especially the effective functional dependence of the matrix elements on $g_A$.
 
%\vspace*{-4mm}
\acknowledgments
%\vspace*{-2mm}

%{\em Acknowledgments.} 
The work of E.L.\  is supported by the Italian Istituto Nazionale di Fisica Nucleare (INFN) and Ministero dell'Istruzione, dell'Universit\`a e della Ricerca (MIUR) through the ``Astroparticle Physics'' 
research projects. The work of and A.M.R.\ is supported by MIUR.
F.\v{S}.\ acknowledges support by the VEGA Grant Agency of the Slovak Republic under the contract N.~1/0876/12.

\end{document}